# Studying the $\Xi(2030)$ as a predominantly $\bar{K}^*\Sigma$ molecular state

Jing-wen Feng and Cai Cheng[*]

*School of Physics and Electronic Engineering, Sichuan Normal University, Chengdu 610101, China*

Yin Huang[†]

*School of Physical Science and Technology, Southwest Jiaotong University, Chengdu 610031, China*

Since its discovery in 1977, the spin parity of $\Xi(2030)$ has not been fully determined experimentally. The latest Particle Data Group listing suggests it may be a baryon with $J = 5/2$. Therefore, studying the mass spectrum and decay properties of $\Xi(2030)$ has become a current hot topic to definitively establish its spin parity. As the three-quark model fails to explain $\Xi(2030)$, we previously proposed it may be a molecule primarily composed of $\bar{K}^*\Sigma$ with $J^P = 5/2^+$, based on its mass spectrum study. To verify its molecular state interpretation, this work proposes studying the strong decays of $\Xi(2030)$ assuming it is a $P$-wave $J^P = 5/2^+$ meson-baryon molecule predominantly composed of $\bar{K}^*\Sigma$. We calculated all experimentally measured two-body and three-body final state decay widths of $\Xi(2030)$, including $\Xi(2030) \to \bar{K}\Lambda, \bar{K}\Sigma, \pi\Xi, \pi\Xi^*$, and $\Xi(2030) \to \pi\pi\Xi, \pi\bar{K}\Sigma, \pi\bar{K}\Lambda$. The results indicate that both the total decay width and partial decay widths agree well with experimental values within the error margins. This supports that $\Xi(2030)$ is a molecule with spin-parity $J^P = 5/2^+$, predominantly composed of $\bar{K}^*\Sigma$. Compared to the experimental central values, our results are slightly smaller, which suggests that $\Xi(2030)$ may contain additional components besides meson-baryon molecular components, such as three-quark structures.

## I. INTRODUCTION

Currently, the utilization of the hadronic molecular state concept is a prominent subject in particle physics, aiming to elucidate and predict exotic states that exhibit a more intricate internal structure than conventional quark states [1–3]. This line of research is primarily motivated by the fact that hadron molecular states manifest as actual entities in nature, rather than solely existing within theoretical models. An exemplar of such a molecular state is the deuteron, a weakly bound state resulting from the binding of a neutron and a proton through the nuclear force. And the nuclear force can bind more neutrons and protons together to form heavier molecular states, such as atomic nuclei $^{208}$Pb. Another well-known candidate for a molecular state is the $\Lambda(1405)$, which is widely believed to have a dominant $\bar{K}N$ component [4–9]. In particular, results from lattice-QCD simulations [6,7] strongly support the interpretation of $\Lambda(1405)$ as a molecular state of $\bar{K}N$ and its coupled channel. Although the nature of the $X(3872)$ cannot be definitively determined as a $\bar{D}D^*$ molecule, its discovery [10] marks the beginning of a new era in the exploration of molecular states. Furthermore, the LHCb Collaboration has reported several hidden-charm pentaquark states [11–14] that exhibit characteristics aligning with $\Sigma_c\bar{D}^{(*)}$ molecules [15–24]. This groundbreaking finding has provided significant impetus for the exploration of molecular states.

It is worth noting that the molecules discussed above are just a few of the well-known candidates for molecular states. There are many molecular states containing other components, especially those with a significant presence of strange quarks. Indeed, there are predictions suggesting that the $f_0(980)$ and $a_0(980)$ are composed of a $\bar{K}K$ molecular component [25,26]. This molecular structure involving $\bar{K}K$ can provide a natural explanation for the significant isospin violations observed in the $\eta(1405/1475) \to \pi\pi\pi$ reaction [27]. The interaction between the $K$ meson and its excited state, $\bar{K}^*$, can form a bound state corresponding to the $f_1(1285)$ resonance [28–33], which is supported by experimental evidence from the BESIII experiment [34]. There exist strong attractions between $K^*$ and $\bar{K}^*$, and the $f_0(1710)$ can be interpreted as a $K^*\bar{K}^*$ molecule [35,36]. Several analyses of the $\phi$ and $K\Lambda(1520)$ photoproduction data suggest the existence of two possible hidden-strangeness molecular states [37–43], which could be associated with the nucleon resonances $N(2100)$ and $N(1875)$, respectively. Indeed, the $\Sigma^*K$ and $\Sigma K^*$ interactions provide a good understanding of these two states [44,45]. The spectroscopy and decay widths of the $\Xi(1620)$ and $\Xi(1690)$ can be well explained by considering them as $\bar{K}\Lambda - \bar{K}\Sigma$ molecular states [46,47]. Additionally, the interaction between the $\bar{K}^*$ and $\Sigma$ gives rise to a $P$-wave molecular state associated with the $\Xi(2030)$ [48]. The presence of a $\bar{K}\Xi(1530)$ hadronic molecular component is crucial in understanding the observed $\Omega(2120)$ [49–53]. The molecular states with $\bar{K}^{(*)}\Xi^{(*)}$ or $K^{(*)}\Omega$ components are also discussed in Refs. [54,55].

---

[*]Contact author: ccheng@sicnu.edu.cn
[†]Contact author: huangy2019@swjtu.edu.cn

In addition to the possible existence of molecules with a strange quark in the light flavor sector, it has been discovered that hadrons containing heavy quarks, such as $c$ or $b$, have the ability to interact with $\bar{K}^{(*)}/K^{(*)}$ mesons, leading to the formation of bound states [56]. These bound states can be found in the Particle Data Group [57]. Notably, two well-known charmed-strange mesons, namely $D_{s0}(2317)$ and $D_{s1}(2460)$, can be regarded as hadronic molecules of $DK$ and $D^*K$, respectively. This interpretation arises from the fact that their masses are approximately 160 and 70 MeV lower than the predicted values based on the quark model [1], and they are about 40 MeV below the thresholds for $DK$ and $D^*K$, respectively. The molecular assignments for $D_{s0}(2317)$ and $D_{s1}(2460)$ have been confirmed by Ref. [58]. Furthermore, the authors in Ref. [58] have identified several additional bound states composed of $D\bar{K}^*$, $D^*\bar{K}^*$, $\bar{K}^{(*)}B^{(*)}$, and $K^{(*)}B^{(*)}$ components, respectively. Among them, the $D^*\bar{K}^*$ molecule corresponds to the experimentally observed $X_0(2900)$.

We can find that the experimental information on the kaonic molecules is abundant. However, the majority of these states are in the $S$-wave, except for the $\Xi(2030)$, which can be identified as a $P$-wave molecular state with mainly the $\bar{K}^*\Sigma$ component [48]. In a similar vein, when considering other components in the molecule, there are only several candidates for the $P$-wave molecule, such as $D_s\bar{D}_{s0}(2370)$ [59] and $\bar{B}^{(*)}N$ [60], which correspond to the experimentally observed states $Y(4274)$ and $\Lambda_b(6146, 6152)$, respectively. Of course, there are theoretical predictions for the existence of the $P$-wave meson-baryon molecule, which include $\bar{D}^{(*)}\Lambda_c$ and $\bar{D}^{(*)}\Sigma_c^{(*)}$ components [61].

One of the main factors contributing to the scarcity of $P$-wave molecules is the repulsive centrifugal potential present in the interaction between hadrons at the $P$-wave, which greatly inhibits the formation of bound states. Interestingly, the deuteron includes an undeniable $D$-wave $NN$ component, prompting further exploration of possible candidates for high wave molecules, especially in the $P$-wave. A possible candidate for a $P$-wave molecular state is the $\Xi(2030)$, discovered as early as 1977 [62]. Currently, the spin parity of the state is not fully determined, with the most likely spin being $J = \frac{5}{2}$ [57]. This excludes the $J^P = \frac{7}{2}^+$ assignment predicted by the constituent quark model [63]. Furthermore, theoretical calculations [64,65] indicate that if the state is a conventional three-quark state with spin-parity $J^P = \frac{5}{2}^+$, its strong decay width does not match experimental observations. Therefore, the authors revisit the idea in the literature [48], proposing that it might be a $P$-wave molecular state with a dominate $\bar{K}^*\Sigma$ component (almost 100%). However, a basic analysis of its mass spectrum alone cannot conclusively determine its molecular state structure, as it may interact with quark states. If the molecular state is predominant, we can also derive its mass spectrum. Another established method involves examining its strong decay width, combined with mass spectrum results, to aid in determining its molecular state structure.

In this study, we investigate the possible strong decay width of $\Xi(2030)$ by treating it as a molecular state with a nearly 100% $\bar{K}^*\Sigma$ component. A definitive understanding of the internal structure of $\Xi(2030)$ can be achieved by comparing the obtained decay width with experimental data. This paper is organized as follows. In Sec. II, we will present the theoretical formalism. In Sec. III, the numerical result will be given, followed by discussions and conclusions in the last section.

## II. FORMALISM AND INGREDIENTS

In this work, we discuss the possibility that $\Xi(2030)$ is a $P$-wave molecular state with a predominant component of $\bar{K}^*\Sigma$ by studying its strong decay properties. The Feynman diagrams considered in this study are depicted in Fig. 1. These diagrams encompass $t$-channel exchanges involving $\pi, \rho, \omega$, and $K$ mesons for the transition from $\Xi(2030)$ to $\bar{K}\Lambda$, $\bar{K}\Sigma$, and $\pi\Xi^{(*)}$ two-body final state reactions. Additionally, the $t$-channel $\bar{K}$ meson exchange is considered for three-body final states, specifically in the reactions $\Xi(2030) \to \pi\pi\Xi$, $\pi\bar{K}\Lambda$, and $\pi\bar{K}\Sigma$. Furthermore, a tree diagram is included to represent the $\Xi(2030) \to \pi\bar{K}\Sigma$ reaction.

To calculate the diagrams depicted in Fig. 1, it is necessary to determine the effective Lagrangian densities corresponding to the relevant interaction vertices. In the case of $\Xi(2030)\Sigma\bar{K}^*$ coupling, we adopt the Lagrangian densities employed in Ref. [66]

$$\mathcal{L}_{\Xi(2030)\Sigma\bar{K}^*}^{J^P=5/2^+} = g_{\Xi(2030)\Sigma\bar{K}^*} \int d^4y \Phi(y^2)\bar{\Sigma}(x+\omega_{\bar{K}^*}y) \\ \times \partial_\mu \bar{K}_\nu^*(x-\omega_\Sigma y)\Xi(2030)^{\mu\nu} + \text{H.c.}, \quad (1)$$

where $\omega_{\bar{K}^*} = m_{\bar{K}^*}/(m_{\bar{K}^*}+m_\Sigma)$ and $\omega_\Sigma = m_\Sigma/(m_{\bar{K}^*}+m_\Sigma)$. The $\Phi(y^2)$ is an effective correlation function. Its physical meaning is to describe the spatial distribution of components within molecules. At the same time, it also serves to prevent ultraviolet divergences in Feynman diagrams during loop diagram calculations. Its form is not fixed, but one of main properties is that it must vanish fast in the ultraviolet region. We adopt the commonly used forms in hadronic molecular decays, and the Fourier transform can be written as

$$\Phi(p_E^2/\Lambda^2) \doteq \exp(-p_E^2/\Lambda^2), \tag{2}$$

where $p_E$ is the Euclidean Jacobi momentum and $\Lambda$ is a free size parameter, typically about 1 GeV, which varies in different systems.

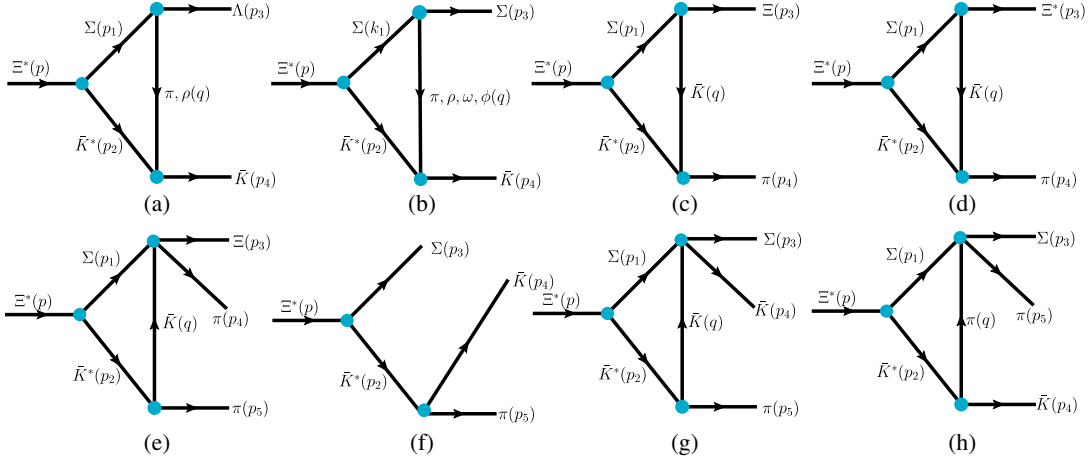

FIG. 1. Feynman diagrams for the $\Xi(2030) \to \bar{K}\Lambda$, $\bar{K}\Sigma$, $\pi\Xi^{(*)}$, $\pi\pi\Xi$, $\pi\bar{K}\Lambda$, and $\pi\bar{K}\Sigma$ reactions. We also show the definition of the kinematics ($p_1$, $p_2$, $p_3$, $p_4$, $p_5$) that we use in the present calculation.

The coupling constant $g_{\Xi(2030)\Sigma\bar{K}^*}$ in Eq. (1) can be established through the compositeness condition [67,68]. This condition stipulates that the renormalization constant of the hadronic molecular wave function must be zero,

$$Z_{\Xi(2030)} = 1 - \frac{\partial \Sigma^T_{\Xi(2030)}(p)}{\partial p^2}\bigg|_{p^2 = m^2_{\Xi(2030)}} = 0, \tag{3}$$

where the transverse part, denoted as $\Sigma^T_{\Xi(2030)}(p)$, of the mass operator $\Sigma^{\mu\nu\alpha\beta}_{\Xi(2030)}(p)$ can be obtained through the following relationships:

$$\Sigma^{\mu\nu\alpha\beta}_{\Xi(2030)}(p) = \frac{1}{2}(g_\perp^{\mu\alpha} g_\perp^{\nu\beta} + g_\perp^{\mu\beta} g_\perp^{\nu\alpha}) \Sigma^T_{\Xi(2030)}(p) + \cdots, \tag{4}$$

with $g_\perp^{\mu\alpha} = g^{\mu\alpha} - p^\mu p^\nu/p^2$. Utilizing the effective Lagrangian shown in Eq. (1), the required mass operator $\Sigma^{\mu\nu\alpha\beta}_{\Xi(2030)}(p)$ can be calculated based on Fig. 2. Its detailed expression is computed as follows:

$$\Sigma^{\mu\nu\alpha\beta}_{\Xi(2030)}(p) = g^2_{\Xi(2030)\Sigma\bar{K}^*} \int \frac{d^4q}{(2\pi)^4 i} \Phi^2(q - \omega_{\bar{K}^*} p)$$
$$\times \frac{1}{\slashed{q} - m_\Sigma}(p-q)^\alpha(p-q)^\mu$$
$$\times \frac{-g^{\beta\nu} + (p-q)^\beta(p-q)^\nu/m^2_{\bar{K}^*}}{(p-q)^2 - m^2_{\bar{K}^*}}, \tag{5}$$

where the $p$ and $q$ are the four-momentum of the $\Xi(2030)$ and $\Sigma$, respectively. $m_{\bar{K}^*}$ and $m_\Sigma$ are the masses of the meson $\bar{K}^*$ and baryon $\Sigma$, respectively. Then, we can obtain the coupling constant of the molecule $\Xi(2030)$ to its components

$$\frac{1}{g^2_{\Xi(2030)\Sigma\bar{K}^*}} = \int d\alpha \int d\beta - \sum_i \frac{\Lambda^4}{\mathcal{C}_i \pi^2 \mathcal{Z}^4} \left[ \frac{\mathcal{F}_i(m_{\bar{K}^*}, \omega_{\bar{K}^*}, m_\Sigma)}{2\mathcal{Z}} \right.$$
$$\left. + \mathcal{H}_i(m_\Sigma) \mathcal{F}'_i(m_{\bar{K}^*}, \omega_{\bar{K}^*}, m_\Sigma) \right], \tag{6}$$

with

$$\mathcal{F}(x, y, z) = \frac{1}{x^2} \exp\left\{-\frac{1}{\Lambda^2}\left[\alpha y^2 + \beta(x^2 - m^2) - 2m^2 z^2 + \frac{m^2(\beta + 2z)^2}{\mathcal{Z}}\right]\right\},$$
$$\mathcal{H}(y) = y - \frac{m}{\mathcal{Z}}, \tag{7}$$

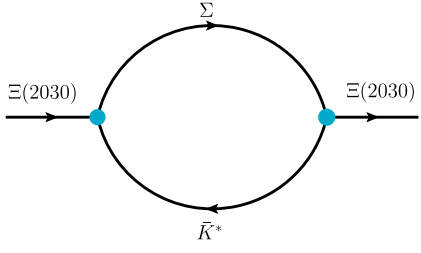

FIG. 2. The mass operator of $\Xi(2030)$.

where $i = \bar{K}^0\Sigma^0$ and $i = \bar{K}^-\Sigma^+$, respectively, and $\mathcal{C}_{\bar{K}^{*0}\Sigma^0} = 256$ and $\mathcal{C}_{\bar{K}^{*-}\Sigma^+} = 192$. $\mathcal{Z} = 2 + \alpha + \beta$ with $\alpha$ and $\beta$ as parameters, which will be integrated from 0 to infinity. $m$ is the mass of the baryon $\Xi(2030)$, and $\mathcal{F}'_i$ is the derivative with respect to $m$.

To calculate the diagrams depicted in Fig. 1, it is also necessary to determine the effective Lagrangian densities for the interaction vertex involving the coupling of a meson to the baryon octet [4,46,69]

$$\mathcal{L}_{VBB} = -g\{\langle \bar{B}\gamma_\mu[V^\mu, B]\rangle + \langle \bar{B}\gamma_\mu B\rangle\langle V^\mu\rangle + \frac{1}{4M}(F\langle \bar{B}\sigma_{\mu\nu}[V^{\mu\nu}, B]\rangle + D\langle \bar{B}\sigma_{\mu\nu}\{V^{\mu\nu}, B\}\rangle)\}, \quad (8)$$

$$\mathcal{L}_{PBB} = \frac{F_1}{2}\langle \bar{B}\gamma_\mu\gamma_5[u^\mu, B]\rangle + \frac{D_1}{2}\langle \bar{B}\gamma_\mu\gamma_5\{u^\mu, B\}\rangle, \quad (9)$$

$$\mathcal{L}_{PBPB} = \frac{i}{4f^2}\langle \bar{B}\gamma^\mu[(P\partial_\mu P - \partial_\mu PP)B - B(P\partial_\mu P - \partial_\mu PP)]\rangle, \quad (10)$$

where the $M$ represent the masses of the baryon and $\langle \cdots \rangle$ refers to an $SU(3)$ trace. $\{A, B\} = AB + BA$ and $[A, B] = AB - BA$. The constants $D = 2.4$ and $F = 0.82$ were found to reproduce well the magnetic moments of the baryons [70]. $F_1 = 0.51$, $D_1 = 0.75$ [46,69], and at the lowest order $u^\mu = -\sqrt{2}\partial^\mu P/f$ with $f = 93$ MeV, The tensor field of the vector mesons $V^{\mu\nu} = \partial^\mu V^\nu - \partial^\nu V^\mu$ and $\sigma^{\mu\nu} = \frac{i}{2}(\gamma^\mu\gamma^\nu - \gamma^\nu\gamma^\mu)$. $B$, $V^\mu$, and $P$ are the SU(3) baryon octet, vector meson, and pseudoscalar matrices, respectively,

$$B = \begin{pmatrix} \frac{1}{\sqrt{2}}\Sigma^0 + \frac{1}{\sqrt{6}}\Lambda & \Sigma^+ & p \\ \Sigma^- & -\frac{1}{\sqrt{2}}\Sigma^0 + \frac{1}{\sqrt{6}}\Lambda & n \\ \Xi^- & \Xi^0 & -\frac{2}{\sqrt{6}}\Lambda \end{pmatrix}, \quad (11)$$

$$V_\mu = \begin{pmatrix} \frac{1}{\sqrt{2}}(\rho^0 + \omega) & \rho^+ & K^{*+} \\ \rho^- & \frac{1}{\sqrt{2}}(-\rho^0 + \omega) & K^{*0} \\ K^{*-} & \bar{K}^{*0} & \phi \end{pmatrix}_\mu, \quad (12)$$

$$P = \begin{pmatrix} \frac{1}{\sqrt{2}}\pi^0 + \frac{1}{\sqrt{6}}\eta & \pi^+ & K^+ \\ \pi^- & -\frac{1}{\sqrt{2}}\pi^0 + \frac{1}{\sqrt{6}}\eta & K^0 \\ K^- & \bar{K}^0 & -\frac{2}{\sqrt{6}}\eta \end{pmatrix}. \quad (13)$$

The Lagrangians involving the interaction of the coupling of the vector meson to the pseudoscalar meson vertex $VVP$ and $VPP$ are given by [71–73]

$$\mathcal{L}_{VVP} = \frac{G'}{\sqrt{2}}\epsilon^{\mu\nu\alpha\beta}\langle\partial_\mu V_\nu \partial_\alpha V_\beta P\rangle, \quad (14)$$

$$\mathcal{L}_{VPP} = -ig\langle[P, \partial_\mu P]V^\mu\rangle, \quad (15)$$

where $\epsilon^{\mu\nu\alpha\beta}$ is the Levi-Civita tensor with $\epsilon^{0123} = 1$. The coupling constant $G' = \frac{3g'^2}{4\pi^2 f}$ with $g' = -\frac{G_V m_\rho}{\sqrt{2}f^2}$, $G_V = 55$ MeV, and $f = 93$ MeV. The coupling constant $g$ can be fixed from the strong decay width of $K^* \to K\pi$. With the help of the Lagrangian shown in Eq. (15), the two-body decay width $\Gamma(K^{*+} \to K^0\pi^+)$ is related to $g$ as

$$\Gamma(K^{*+} \to K^0\pi^+) = \frac{g^2}{6\pi m_{K^{*+}}^2}\mathcal{P}_{\pi K^*}^3 = \frac{2}{3}\Gamma_{K^{*+}}, \quad (16)$$

where the $\mathcal{P}_{\pi K^*}$ is the three momentum of the $\pi$ in the rest frame of the $K^*$. Using the experimental strong decay width $\Gamma_{K^{*+}} = 50.3 \pm 0.8$ MeV and the masses of the particles [57], we obtain $g = 4.64$.

To evaluate the decay amplitudes, we also need the following effective Lagrangians to calculate the relevant vertices [74]:

$$\mathcal{L}_{\Xi^*\Sigma K} = \frac{f_{\Xi^*\Sigma K}}{m_K}\partial^\nu \bar{K}\bar{\Xi}_\nu^*\gamma_5\vec{\tau}\cdot\vec{\Sigma} + \text{H.c.}, \quad (17)$$

where the coupling constant $f_{\Xi^*\Sigma K} = 3.22$ and $m_K$ is the mass of the $K$ meson.

Putting all the pieces together, we obtain the following strong decay amplitudes:

$$\mathcal{M}_a^{\{\pi^0,\pi^+\}} = (i)^3 \frac{D_1 g g_{\Xi(2030)\Sigma\bar{K}^*}}{3\sqrt{2}f}\left\{1, \frac{2}{\sqrt{3}}\right\}\int \frac{d^4q}{(2\pi)^4 i}\Phi$$
$$\times [(p_1\omega_{\bar{K}^*} - p_2\omega_\Sigma)_E^2]\bar{u}(p_3)\slashed{q}\gamma_5\frac{\slashed{p}_1 + m_\Sigma}{p_1^2 - m_\Sigma^2}u_{\nu\sigma}(k)$$
$$\times p_2^\nu(p_4^\alpha + q^\alpha)\frac{-g^{\sigma\alpha} + p_2^\sigma p_2^\alpha/m_{\bar{K}^*}^2}{p_2^2 - m_{\bar{K}^*}^2}\frac{1}{q^2 - m_\pi^2},$$
$$(18)$$

$$\mathcal{M}_a^{\{\rho^0,\rho^+\}} = i(i)^3 \frac{DG'g_{\Xi(2030)\Sigma\bar{K}^*}}{12\sqrt{2}m_\Sigma} \left\{1, \frac{2}{\sqrt{3}}\right\} \int \frac{d^4q}{(2\pi)^4 i} \Phi[(p_1\omega_{\bar{K}^*} - p_2\omega_\Sigma)_E^2] \bar{u}(p_3)(\slashed{q}\gamma_\eta - \gamma_\eta\slashed{q})$$
$$\times \frac{\slashed{p}_1 + m_\Sigma}{p_1^2 - m_\Sigma^2} u_{\theta\lambda}(k)\epsilon^{\sigma\tau\alpha\beta} p_{2\sigma} q_\alpha p_2^\theta \frac{-g^{\lambda\beta} + p_2^\lambda p_2^\beta/m_{\bar{K}^*}^2}{p_2^2 - m_{\bar{K}^*}^2} \frac{-g^{\tau\eta} + q^\tau q^\eta/m_\rho^2}{q^2 - m_\rho^2}, \quad (19)$$

$$\mathcal{M}_b^{\{\pi^0,\pi^+\}} = (i)^3 \frac{F_1 g g_{\Xi(2030)\Sigma\bar{K}^*}}{f} \left\{0, -\sqrt{\frac{2}{3}}\right\} \int \frac{d^4q}{(2\pi)^4 i} \Phi[(p_1\omega_{\bar{K}^*} - p_2\omega_\Sigma)_E^2] \bar{u}(p_3)\slashed{q}\gamma_5 \frac{\slashed{p}_1 + m_\Sigma}{p_1^2 - m_\Sigma^2} u_{\nu\sigma}(k)$$
$$\times p_2^\nu (p_4^\alpha + q^\alpha) \frac{-g^{\sigma\alpha} + p_2^\sigma p_2^\alpha/m_{\bar{K}^*}^2}{p_2^2 - m_{\bar{K}^*}^2} \frac{1}{q^2 - m_\pi^2}, \quad (20)$$

$$\mathcal{M}_b^\eta = (i)^3 \frac{D_1 g g_{\Xi(2030)\Sigma\bar{K}^*}}{3\sqrt{6}f} \int \frac{d^4q}{(2\pi)^4 i} \Phi[(p_1\omega_{\bar{K}^*} - p_2\omega_\Sigma)_E^2] \bar{u}(p_3)\slashed{q}\gamma_5 \frac{\slashed{p}_1 + m_\Sigma}{p_1^2 - m_\Sigma^2} u_{\nu\sigma}(k) p_2^\nu (p_4^\alpha + q^\alpha)$$
$$\times \frac{-g^{\sigma\alpha} + p_2^\sigma p_2^\alpha/m_{\bar{K}^*}^2}{p_2^2 - m_{\bar{K}^*}^2} \frac{1}{q^2 - m_\eta^2}, \quad (21)$$

$$\mathcal{M}_b^{\{\rho^0,\rho^+\}} = -i(i)^3 G'g_{\Xi(2030)\Sigma\bar{K}^*}\{0, \sqrt{2}\} \int \frac{d^4q}{(2\pi)^4 i} \Phi[(p_1\omega_{\bar{K}^*} - p_2\omega_\Sigma)_E^2] \bar{u}(p_3)\epsilon^{\alpha\beta\theta\lambda} q_\alpha p_{2\theta}$$
$$\times \left[g\gamma_\eta + \frac{F}{4}(\slashed{q}\gamma_\eta - \gamma_\eta\slashed{q})\right] \frac{\slashed{p}_1 + m_\Sigma}{p_1^2 - m_\Sigma^2} u_{\sigma\tau}(k) p_2^\sigma \frac{-g^{\tau\lambda} + p_2^\tau p_2^\lambda/m_{\bar{K}^*}^2}{p_2^2 - m_{\bar{K}^*}^2} \frac{-g^{\eta\beta} + q^\eta q^\beta/m_{\rho^-}^2}{q^2 - m_\rho^2}, \quad (22)$$

$$\mathcal{M}_b^{\{\omega,\phi\}} = -i(i)^3 \frac{G'g_{\Xi(2030)\Sigma\bar{K}^*}}{\sqrt{3}} \left\{\sqrt{3}, \sqrt{\frac{3}{2}}\right\} \int \frac{d^4q}{(2\pi)^4 i} \Phi[(p_1\omega_{\bar{K}^*} - p_2\omega_\Sigma)_E^2] \bar{u}(p_3)\left[g\gamma_\eta + \left\{\frac{D}{4}(\slashed{q}\gamma_\eta - \gamma_\eta\slashed{q}), 0\right\}\right]$$
$$\times \frac{\slashed{p}_1 + m_\Sigma}{p_1^2 - m_\Sigma^2} u_{\sigma\tau}(k) p_2^\sigma \epsilon^{\alpha\beta\theta\lambda} q_\alpha p_{2\theta} \frac{-g^{\tau\lambda} + p_2^\tau p_2^\lambda/m_{\bar{K}^*}^2}{p_2^2 - m_{\bar{K}^*}^2} \frac{-g^{\eta\beta} + q^\eta q^\beta/m_{\omega/\phi}^2}{q^2 - m_{\omega/\phi}^2}, \quad (23)$$

$$\mathcal{M}_c^{\{\bar{K}^0,K^-\}} = -(i)^3 \frac{(D_1 + F_1)g g_{\Xi(2030)\Sigma\bar{K}^*}}{2\sqrt{6}f} \{1, -\sqrt{2}\} \int \frac{d^4q}{(2\pi)^4 i} \Phi[(p_1\omega_{\bar{K}^*} - p_2\omega_\Sigma)_E^2] \bar{u}(p_3)\slashed{q}\gamma_5 \frac{\slashed{p}_1 + m_\Sigma}{p_1^2 - m_\Sigma^2} u_{\nu\sigma}(k)$$
$$\times p_2^\nu (p_4^\alpha + q^\alpha) \frac{-g^{\sigma\alpha} + p_2^\sigma p_2^\alpha/m_{\bar{K}^*}^2}{p_2^2 - m_{\bar{K}^*}^2} \frac{1}{q^2 - m_{\bar{K}}^2}, \quad (24)$$

$$\mathcal{M}_d^{\{\bar{K}^0,K^-\}} = -(i)^3 \frac{f_{\Xi^*\Sigma K} g g_{\Xi(2030)\Sigma\bar{K}^*}}{\sqrt{6}m_K} \{1, -\sqrt{2}\} \int \frac{d^4q}{(2\pi)^4 i} \Phi[(p_1\omega_{\bar{K}^*} - p_2\omega_\Sigma)_E^2] q \cdot \bar{u}(p_3)\gamma_5 \frac{\slashed{p}_1 + m_\Sigma}{p_1^2 - m_\Sigma^2} u_{\nu\sigma}(k)$$
$$\times p_2^\nu (p_4^\alpha + q^\alpha) \frac{-g^{\sigma\alpha} + p_2^\sigma p_2^\alpha/m_{\bar{K}^*}^2}{p_2^2 - m_{\bar{K}^*}^2} \frac{1}{q^2 - m_{\bar{K}}^2}, \quad (25)$$

$$\mathcal{M}_e^{\{\bar{K}^0,K^-\}} = -i(i)^3 \frac{g g_{\Xi(2030)\Sigma\bar{K}^*}}{8\sqrt{6}f^2} \int \frac{d^4q}{(2\pi)^4 i} \Phi[(p_1\omega_{\bar{K}^*} - p_2\omega_\Sigma)_E^2] \bar{u}(p_3)(\slashed{p}_4 + \slashed{q}) \frac{\slashed{p}_1 + m_\Sigma}{p_1^2 - m_\Sigma^2} u_{\nu\sigma}(k) p_2^\sigma \frac{-g^{\sigma\tau} + p_2^\sigma p_2^\tau/m_{\bar{K}^*}^2}{p_2^2 - m_{\bar{K}^*}^2}$$
$$\times (p_5^\tau - q^\tau) \frac{1}{q^2 - m_{\bar{K}}^2} \{1, -\sqrt{2}\}, \quad (26)$$

$$\mathcal{M}_f^{\{\bar{K}^0,K^-\}} = -i(i)^3 \frac{g g_{\Xi(2030)\Sigma\bar{K}^*}}{2\sqrt{6}} \Phi[(p_3\omega_{\bar{K}^*} - p_2\omega_\Sigma)_E^2] \bar{u}(p_3) u_{\nu\alpha}(k) p_2^\alpha \frac{-g^{\mu\nu} + p_2^\mu p_2^\nu/m_{\bar{K}^*}^2}{p_2^2 - m_{\bar{K}^*}^2} (p_5^\mu - p_4^\mu)\{1, -\sqrt{2}\}, \quad (27)$$

$$\mathcal{M}_g^{\{\bar{K}^0,K^-\}} = i(i)^3 \frac{gg_{\Xi(2030)\Sigma\bar{K}^*}}{2\sqrt{6}f^2} \int \frac{d^4q}{(2\pi)^4 i} \Phi[(p_1\omega_{\bar{K}^*} - p_2\omega_\Sigma)_E^2] \bar{u}(p_3)(\slashed{p}_4 + \slashed{q}) \frac{\slashed{p}_1 + m_\Sigma}{p_1^2 - m_\Sigma^2} u_{\nu\sigma}(k) p_2^\nu \frac{-g^{\sigma\tau} + p_2^\sigma p_2^\tau/m_{\bar{K}^*}^2}{p_2^2 - m_{\bar{K}^*}^2}$$
$$\times (p_5^\tau - q^\tau) \frac{1}{q^2 - m_{\bar{K}}^2} \{0, 1\}, \tag{28}$$

$$\mathcal{M}_h^{\{\bar{K}^0,K^-\}} = i(i)^3 \frac{gg_{\Xi(2030)\Sigma\bar{K}^*}}{\sqrt{6}f^2} \int \frac{d^4q}{(2\pi)^4 i} \Phi[(p_1\omega_{\bar{K}^*} - p_2\omega_\Sigma)_E^2] \bar{u}(p_3)(\slashed{p}_5 + \slashed{q}) \frac{\slashed{p}_1 + m_\Sigma}{p_1^2 - m_\Sigma^2} u_{\nu\sigma}(k) p_2^\nu \frac{-g^{\sigma\tau} + p_2^\sigma p_2^\tau/m_{\bar{K}^*}^2}{p_2^2 - m_{\bar{K}^*}^2}$$
$$\times (p_4^\tau - q^\tau) \frac{1}{q^2 - m_{\bar{K}}^2} \{0, \sqrt{2}\}, \tag{29}$$

where $\{A, B\}$ are for $\{\pi^0, \pi^+\}$ exchange, respectively.

Once the amplitudes are calculated, we can obtain the partial decay widths. The corresponding formulas for the two-body and three-body decay widths can be found in Ref. [57]; we do not provide the detailed forms here.

## III. RESULTS AND DISCUSSIONS

In this work, we study the strong decay of $\Xi(2030)$ as a $P$-wave molecule with $J^P = 5/2^+$ using the effective Lagrangian approach. If the estimated strong decay width matches well with experimental observations, we can validate the molecular interpretation of $\Xi(2030)$, primarily as a $\bar{K}^*\Sigma$ component, as first proposed in Ref. [48]. Currently, the accuracy of our calculations hinges primarily on the parameter $\Lambda$, necessitating a preliminary discussion of its value. As mentioned earlier, $\Lambda$ is usually set to 1 GeV, but we have observed variations in $\Lambda$ ranging from 0.91 to 1.0 GeV through fitting experimental data [75]. It is important to note that this range is determined using the same theoretical framework as applied in this study. Moreover, we have found that many exotic states can be effectively described as molecules with $\Lambda$ values ranging from 0.90 to 1.10 GeV. For further details, we refer the reader to review Refs. [47,76,77] and their references. Therefore, we choose to explore the range $\Lambda = 0.90$–$1.10$ GeV to investigate whether $\Xi(2030)$ can be interpreted as a molecule primarily composed of $\bar{K}^*\Sigma$ molecular components.

Taking into account the $\Lambda$ values adopted in this work, we first discuss the results of the calculated coupling constants. By substituting Eq. (7) into Eq. (6), we determine the dependence of the coupling constants on the parameter $\Lambda$. Integrating parameters $\alpha$ and $\beta$ from 0 to infinity, we present the numerical results of the coupling constants as $\Lambda$ varies from 0.90 to 1.10 GeV in Fig. 3. From the results, it is evident that the coupling constants decrease as $\Lambda$ increases, indicating a high sensitivity to changes in $\Lambda$, with the range of variation decreasing from 58.26 to 49.92. Specifically, at $\Lambda = 1.0$ GeV, the coupling constant is $g_{\Xi(2030)\Sigma\bar{K}^*} = 53.73$.

After determining the coupling constants, we calculate the partial decay widths for the two-body final state transitions $\Xi(2030) \to \bar{K}\Lambda, \bar{K}\Sigma, \pi\Xi, \pi\Xi^*$ and the three-body final state transition $\Xi(2030) \to \pi\pi\Xi, \pi\bar{K}\Sigma$. These results are depicted in Fig. 4 and vary with the parameter $\Lambda = 0.9$–$1.1$ GeV. It is important to note that the partial decay widths in the channels $\Xi(2030) \to \bar{K}^0\Lambda, \bar{K}^0\Sigma^0, \pi^0\Xi^0, \pi^0\Xi^{*0}$ and $\Xi(2030) \to \pi^0\pi^0\Xi^0, \pi^0\bar{K}^0\Sigma^0$ are estimated here; other channels can be determined using isospin symmetry. The sum of these partial decay widths yields the total decay width of $\Xi(2030)$, as also shown in Fig. 4.

From the results shown in Fig. 4, it is evident that the total decay width, depicted by the solid black line, decreases gradually as $\Lambda$ increases. As $\Lambda$ increases from 0.9 GeV to 1.1 GeV, the predicted total decay width decreases from 17.041 MeV to 13.161 MeV, closely matching the experimental value of $20^{+15}_{-5}$ MeV [57], indicated by cyan error bands. This suggests that if the spin parity of the $\Xi(2030)$ is $J^P = 5/2^+$, the assignment as a $P$-wave molecule predominantly composed of a $\bar{K}^*\Sigma$ component for the $\Xi(2030)$ is supported. The results also indicate that the main contribution comes from the $\bar{K}\Sigma$ channel (dotted blue line), with the $\bar{K}\Lambda$ channel (red dashed line) being the second most important, and the other decay channels being much smaller. We list the detailed numerical results in Table I. For ease of comparison, we also provide the experimental results [57] in the last column.

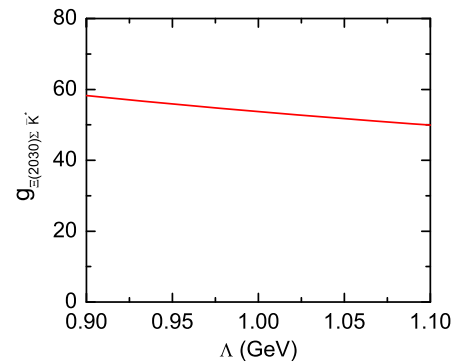

FIG. 3. The coupling constants of the $\Xi(2030)$ assuming it is a $\bar{K}^*\Sigma$ molecule as a function of the parameter $\Lambda$.

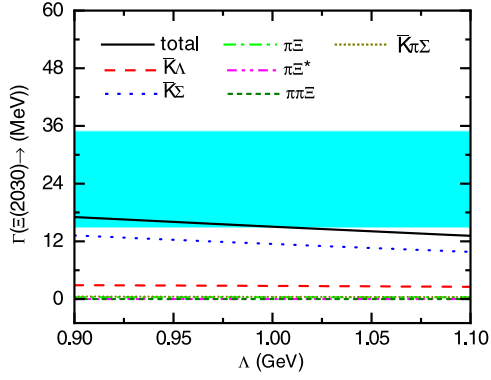

FIG. 4. The decay widths of $\Xi(2030)$ to the final states $\Lambda\bar{K}$, $\Sigma\bar{K}$, $\Xi^*\pi$, $\Xi\pi\pi$, and $\Sigma\bar{K}\pi$, along with the total decay width. The cyan error bands correspond to the experimental values [57].

From Table I, it can be seen that at $\Lambda = 0.9$–1.1 GeV, the numerical results for the two main decay channels, $\Xi(2030) \to \bar{K}\Sigma$ and $\Xi(2030) \to \bar{K}\Lambda$, are 9.816–13.224 MeV and 2.547–2.880 MeV, respectively, contributing 74.58%–77.60% and 16.9%–19.35% to the total decay width. These results are very consistent with the experimentally measured decay widths $\Gamma(\Xi(2030) \to \bar{K}\Sigma) = 12$–28 MeV and $\Gamma(\Xi(2030) \to \bar{K}\Lambda) = 3$–7 MeV, which account for 80% and 20% of the total decay width, respectively [57]. This further supports the reasonableness of considering $\Xi(2030)$ as a molecule primarily composed of $\bar{K}^*\Sigma$ with a spin parity of $J^p = 5/2^+$. Table I also shows that the estimated two-body decay widths of $\Xi(2030)$ into $\Xi\pi$ and $\Xi^*\pi$ are approximately 0.376–0.391 MeV and 0.00579–0.00746 MeV, respectively. In experiments, specific values for these two decay channels have not been provided and are only considered to be small. Our numerical results are similarly small, accounting for 2.294%–2.857% and 0.0438%–0.0440% of the total decay width, respectively, which appears to be consistent with the experimental observations. This means our results also support that the transitions $\Xi(2030) \to \Xi\pi$ and $\Xi(2030) \to \Xi^*\pi$ contribute only a minor portion to the total decay width.

The same conclusion about the small contributions also applies to the $\Xi(2030) \to \pi\pi\Xi$ (not via the intermediate state $\pi\Xi^*$), $\Xi(2030) \to \pi\bar{K}\Lambda$, and $\Xi(2030) \to \pi\bar{K}\Sigma$ reactions observed in experiments, which is also very consistent with the results we have provided (see Table I). For the $\Xi(2030) \to \pi\pi\Xi$ reaction, the decay width is up to 0.00367–0.0120 MeV, which accounts for 0.0279%–0.0704% of total width. That means the transition $\Xi(2030) \to \pi\pi\Xi$ gives a minor contribution. Our calculations also find that the decay width of $\Xi(2030) \to \pi\bar{K}\Sigma$ is 0.409–0.528 MeV, which also accounts for only a small part of the total decay width, being 3.098%–3.107%.

It is noteworthy that the decay $\Xi(2030) \to \pi\bar{K}\Sigma$ includes a tree-level contribution [see Fig. 1(f)], which typically should be dominant and the decay width may even exceed that the main decay channel $\Xi(2030) \to \bar{K}\Sigma$. However, we find that within the parameter range $\Lambda = 0.9$–1.1 GeV, the decay width due to the tree-level contribution is only 0.918–1.185 MeV (see Fig. 5), which is much smaller than the decay width of $\Xi(2030) \to \bar{K}\Sigma$, the main decay channel observed in experiments. A possible explanation for this may be that the phase space for the transition $\Xi(2030) \to \pi\bar{K}\Sigma$ is smaller than that of the $\Xi(2030) \to \bar{K}\Sigma$ reaction. In Fig. 5, we also present the other contributions to the $\Xi(2030) \to \bar{K}\Sigma$ decay. We can observe that the contribution from the $\pi$-meson exchange is smaller than that of the tree-level diagram, and the contribution from the $K$-meson exchange is the smallest. Additionally, the interference effects between these contributions are significant, resulting in the total decay width being smaller than the tree-level decay width.

In Table I, we do not provide the numerical result for the decay width of $\Xi(2030) \to \pi\bar{K}\Lambda$ mainly because the decay of $\Xi(2030)$ to $\pi\bar{K}\Lambda$ via the $\bar{K}^*\Sigma$ molecule is forbidden due to the absence of the $\bar{K}\Sigma \to \bar{K}\Lambda$ and $\pi\Sigma \to \pi\Lambda$ coupling vertices in Eq. (10). However, we find that $\Xi(2030)$ also contains a very small $\bar{K}^*\Lambda$ component [48] that can decay to $\pi\bar{K}\Lambda$. Therefore, we can directly deduce that the decay width of $\Xi(2030)$ to $\pi\bar{K}\Lambda$ via the $\bar{K}^*\Lambda$ component is very small, which is also comparable to experimental results.

TABLE I. Partial decay widths of $\Xi(2030)$ and the total decay width with $\Lambda = 0.9$–1.1 GeV, compared with experimental values from Ref. [57]. The value in parentheses in the last line corresponds to $\Lambda = 1.0$ GeV. All widths are given in units of MeV.

| Model | Our work | Exp. [57] |
| --- | --- | --- |
| $\Xi(2030) \to \bar{K}\Sigma$ | 9.816–13.224 | 12–28 |
| $\to \bar{K}\Lambda$ | 2.547–2.880 | 3–7 |
| $\to \pi\Xi$ | 0.376–0.391 | Small |
| $\to \pi\Xi^*$ | 0.00579–0.00746 | Small |
| $\to \pi\pi\Xi$[not $\pi\Xi^*$] | 0.00367–0.0120 | Small |
| $\to \pi\bar{K}\Sigma$ | 0.409–0.528 | Small |
| $\to \pi\bar{K}\Lambda$ | Small | Small |
| Total | 13.161–17.041 (15.054) | $20^{+15}_{-5}$ |

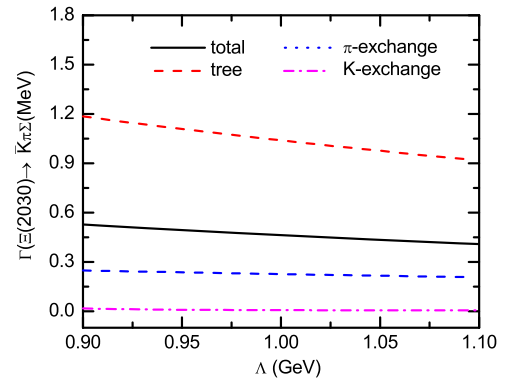

FIG. 5. The decay width of $\Xi(2030) \to \Sigma\bar{K}\pi$ with meson exchange and tree-level contribution.

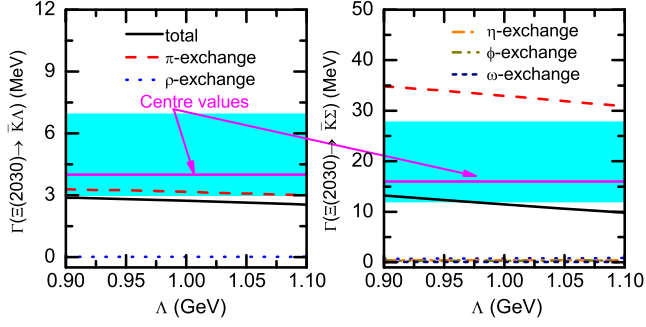

FIG. 6. Partial decay widths of the $\Xi(2030) \to \Lambda \bar{K}$ and $\Xi(2030) \to \Sigma \bar{K}$. The cyan error bands correspond to the experimental values [57].

It is noteworthy that while our results support that $\Xi(2030)$ is a molecule primarily composed of the $\bar{K}^*\Sigma$ component, the calculated total width, particularly with $\Lambda = 1.0$, results in a width of $\Gamma_{\Xi(2030)} = 15.054$ MeV, which is smaller than the experimental central value of $\Gamma_{\Xi(2030)} = 20.0$ MeV. This suggests that $\Xi(2030)$ might not be solely a $P$-wave meson-baryon molecule [48]; it could also include additional three-quark components. A detailed comparison between the theoretical calculations and experimental results for the two main decay channels, $\Xi(2030) \to \Sigma\bar{K}$ and $\Xi(2030) \to \Lambda\bar{K}$, is presented in Fig. 6. We can find that the obtained decay widths for these two channels are smaller than the experimental central values, further supporting the conclusion that a molecular assignment for $\Xi(2030)$ is insufficient.

Figure 6 also indicates that the $\pi$ meson exchange plays a predominant role in the decays $\Xi(2030) \to \Lambda\bar{K}$ and $\Xi(2030) \to \Sigma\bar{K}$, contributing significantly more than the total width. However, contributions from other mesons, including $\rho$, $\omega$, $\eta$, and $\phi$, are almost negligible. Yet, their interference effects are very strong, resulting in a total decay width smaller than that from the $\pi$ meson exchange alone.

## IV. SUMMARY

Inspired by the failure of the quark model to explain $\Xi(2030)$ as a three-quark state, we verify our proposal that $\Xi(2030)$ is a molecule with spin parity $J^P = 5/2^+$ and predominantly composed of $\bar{K}^*\Sigma$ by studying the allowed two-body and three-body strong decays of $\Xi(2030)$. With the $\bar{K}^*\Sigma$ assignment, we calculated the partial decay widths of $\Xi(2030)$ into the two-body final states $\bar{K}\Lambda, \bar{K}\Sigma, \pi\Xi, \pi\Xi^*$, and the three-body final states $\pi\pi\Xi, \pi\bar{K}\Sigma, \pi\bar{K}\Lambda$, some of which occur through hadronic loops. The decay process is described by $t$-channel exchanges of $\pi$, $K$, $\rho$, $\omega$, and $\phi$ mesons, as well as tree-level contributions.

Our numerical results indicate that $\bar{K}\Lambda$ and $\bar{K}\Sigma$ are the two main decay channels of $\Xi(2030)$, with decay widths of $\Gamma(\Xi(2030) \to \bar{K}\Sigma) = 9.816$–$13.224$ MeV and $\Gamma(\Xi(2030) \to \bar{K}\Lambda) = 2.547$–$2.880$ MeV, accounting for 74.58%–77.60% and 16.9%–19.35% of the total decay width, respectively. However, other decay channels provide very small contributions and can be almost neglected. These decay widths result in a total width of $13.161$–$17.041$ MeV. When comparing with the experimental results, we find that both the total decay width and the partial decay widths agree well with the experimental values within the error margins. This supports the hypothesis that $\Xi(2030)$ is a molecule with spin parity $J^P = 5/2^+$, predominantly composed of $\bar{K}^*\Sigma$. However, our results are slightly smaller than the experimental central values, suggesting that $\Xi(2030)$ may also contain additional components, such as three-quark structures, besides the meson-baryon molecular components.

## ACKNOWLEDGMENTS

This work was supported by the National Natural Science Foundation of China under Grant No. 12005177.